# TCP OVER IEEE 802.11

**P. Chenna Reddy**
JNTU College of Engg., Pulivendula, Andhra Pradesh, India – 516390
pcreddy1@rediffmail.com

**ABSTRACT.** IEEE 802.11 is a widely used wireless LAN standard for medium access control. TCP is a prominent transport protocol originally designed for wired networks. TCP treats packet loss as congestion and reduces the data rate. In wireless networks packets are lost not only due to congestion but also due to various other reasons. Hence there is need for making TCP adaptable to wireless networks. Various parameters of TCP and IEEE 802.11 can be set to appropriate values to achieve optimum performance results. In this paper optimum values for various parameters of IEEE 802.11 are determined. Network simulator NS2 is used for simulation.

## Introduction

Number of users using wireless devices for communication has exceeded the number of users using wired devices. Because of low cost and ease of installation wireless networks are preferred over wired networks in Offices, Universities etc. Wireless networks use shared medium for transmission, and the Medium access control sub layer controls access to the shared medium. Various wireless communication standards have evolved to control access to the wireless medium and IEEE 802.11 is the most widely used one.

IEEE 802.11 specification supports two different MAC schemes [C+97], namely Distributed Coordination Function (DCF), and Point Coordination Function (PCF). DCF is used in the contention period. The contention period of operation in which medium access is controlled by DCF is called adhoc mode of WLAN or adhoc network.

173

DCF uses medium access method called Carrier Sense Multiple Access with Collision Avoidance (CSMA/CA). Before sending a frame a node has to sense the channel. If the medium is at least idle for DIFS (Distributed Inter Frame Space) time interval, then node is allowed to send immediately. If the medium is busy, node has to wait till the end of current transmission, after that node waits again for DIFS time period, and then node defers for a random backoff-interval before transmitting.

Every node chooses its own backoff-interval. Node selects a random number of time slots between 0 and backoff-interval, according to the equation, backoff-interval = SlotTime*Random, where Random is a pseudo-random integer value chosen out of uniformly distributed contention window [0, CW], and CW is an integer between contention window minimum (CWMin) and contention window maximum (CWMax). The backoff-interval is decreased by one, for every slot, as long as the channel is sensed idle in that slot. If there are transmissions by other nodes during the slot, then the node freezes its backoff-interval and resumes the count where it left off, after the other transmitting node has completed its frame transmission plus an additional DIFS interval. Transmission commences when the backoff-interval reaches zero. If the transmission is successful, then the CW is set to CWMin, otherwise the value of CW is doubled. The maximum value of CW is CWMax.

IEEE 802.11 uses two retry counts to determine the number of retransmission attempts, Short Retry Count and Long Retry Count. Retry Count is the maximum number of Data transmission attempts that are made before a frame is discarded. Short Retry Count is used for frames whose size is below an RTS-Threshold value. Its default value is 7. Long Retry count, default value is 4, is used for frames whose size is greater than RTS-Threshold value.

**I. Related Work**

TCP performance achieved using Snoop protocol over an IEEE 802.11b platform is evaluated in [G+01]. Exhaustive measurement is carried out over a single hop networks in both good and bad channel conditions. The results obtained show that when wireless channel conditions make packet loss negligible, Snoop TCP doesn't introduce any overhead that results in throughput reduction. However, in a channel with low Signal to Noise ratio, working at 11Mbps rate, a high number of wireless errors produced in long bursts, is observed. TCP is not able to deal properly with these losses and throughput of TCP connection falls below 1Mbps.

The results of an extensive experimental study of the performance of the TCP protocol over wireless multi-hop adhoc networks are presented in [KK05]. The investigations are performed in a real indoor environment over a network of laptops equipped with off-the-shelf IEEE 802.11b wireless cards. The cards were partially covered with copper tape to reduce their range, which enabled creation of manageable topologies. The important result of the study is a recommendation of few TCP and IEEE 802.11 parameters that are best for TCP performance over wireless multi-hop networks.

The effect of congestion and MAC contention on the interaction between TCP and on-demand adhoc routing protocol in the 802.11 ad hoc networks is presented in [NJ05]. It is observed that TCP induces the over-reaction of routing protocol and hurts the quality of end-to-end connection. The reason for deterioration in quality lies in the TCP window mechanism itself. To fix this problem, a fractional window increment scheme for TCP by limiting TCP's aggressiveness is proposed. The proposed scheme dramatically improves TCP performance and network stability in a variety of 802.11 multihop networks.

A model for TCP over 802.11 networks is designed in [KS04] and it is applied to study the effect of ACK thinning due to delayed acknowledgement option of TCP. Significant improvement in performance is observed due to delayed acknowledgements.

TCP performance in multi-hop adhoc networks under static chain topology is presented in [W+04]. Effects of MAC retransmission limit on End-to-End TCP performance are evaluated with different background loads, which indicate that when the network load is light, larger retransmission limit makes TCP throughput more appreciable; when the load is heavy, too large retransmission limit result in drastic performance degradation. In addition, simulation results under dynamic adhoc topology indicate that topology changes of adhoc network itself are not sensitive to MAC retransmission limits.

TCP performance in a stationary multihop wireless network using IEEE 802.11 for channel access control is studied in [F+05]. Given a specific network topology and flow patterns, there exists an optimal window size 'W' at which TCP achieves the highest throughput via maximum spatial reuse of the shared wireless channel. However, TCP grows its window size much larger than W, leading to throughput reduction. As the offered load increases, probability of packet drops due to link contention also increases, and eventually saturates. Unfortunately, link layer drop probability is insufficient to keep the TCP window size around W. Link RED that fine tunes the link layer packet dropping probability to stabilize TCP window size around W is proposed.

## II. Performance of TCP over exiting IEEE 802.11

In this section Optimum values for various IEEE 802.11 parameters are determined. The parameters considered are Short Retry count, CWMin, and CWMax.

### A. Methodology

NS-2.26 is used as network simulator and graphs are generated using Xgraph. Simulation environment consists of 10, 12, and 14 wireless nodes forming a static adhoc network in a flat space of 1500m X 1200m. The distance between the nodes is approximately 200m. Total of four sources generate FTP traffic with packet size 1500 bytes. The topology is dumbell topology with varying number of intermediate nodes as shown in Fig. 1. If the number of intermediate nodes is not mentioned, then number of intermediate nodes is assumed as 6.

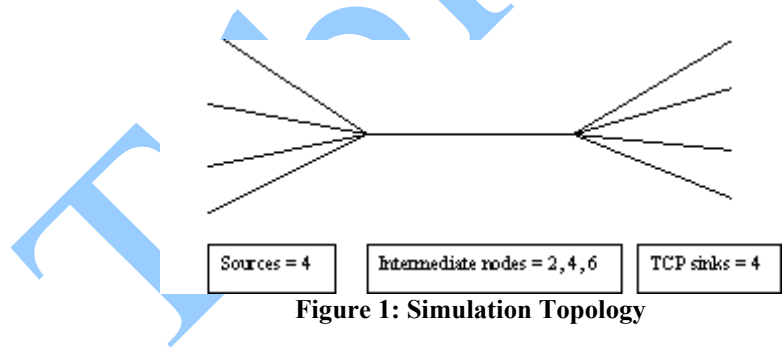

**Figure 1: Simulation Topology**

Each run of the simulation is for 200 seconds. The physical radio characteristics of each mobile node's network interface, such as the antenna gain, transmit power, and receiver sensitivity, were chosen to approximate the Lucent WaveLAN Direct Sequence Spread Spectrum (DSSS) radio. The parameter settings of Lucent WaveLAN Direct Sequence Spread Spectrum are as shown in Table 1. Adhoc network routing protocol used is Dynamic Source Routing.

176

**Table 1: DSSS Parameter settings**

| Parameter | Value |
|---|---|
| Maximum bandwidth | 2 Mbps |
| Frequency | 914MHz |
| Modulation | DSSS/DQPSK |
| Capture Threshold | 10.0 dB |
| Carrier Sense Threshold | 1.559 e -11 dBm |
| Receiver Threshold | 3.562 e -10 dBm |

**B. Metrics**

*Number of packets delivered successfully*: It is the total number of data packets received by the destination.
*Average Delay*: It is the average amount of time a packet takes to go from source to destination. It is an End-to-End delay.
*Number of packets dropped*: It is the total number of data packets dropped.
*MAC Drops*: It is the total number of data packets dropped due to various problems at the MAC layer. Packets dropped due to collisions and packets dropped due to retry count exceeding its limit and other reasons are differentiated.

**C. Simulation Results and Analysis**

**The effect of Short Retry Count**

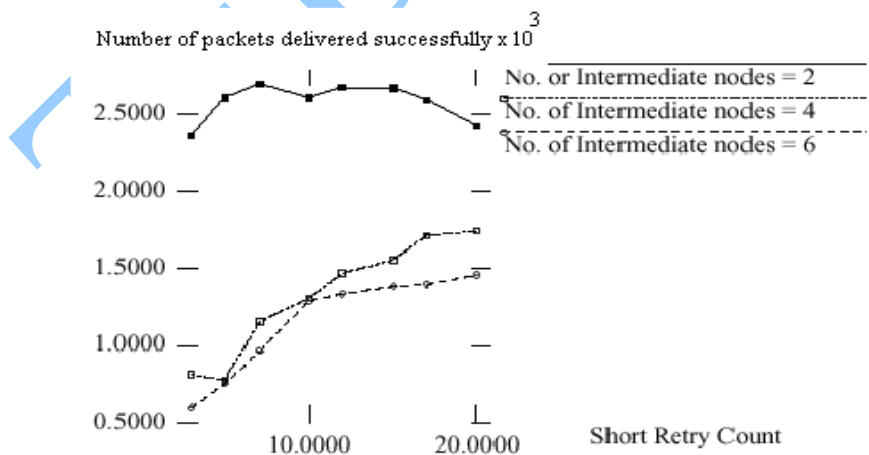

**Figure 2: Variation of Number of packets delivered successfully with Short Retry Count**

177

**Short Explanation**: *The numbers of packets delivered successfully increases with the increase in the value of short retry count. Maximum number of packets is delivered when short retry limit is 20. As the number of intermediate nodes increases it is better to use higher values of short retry limit.*

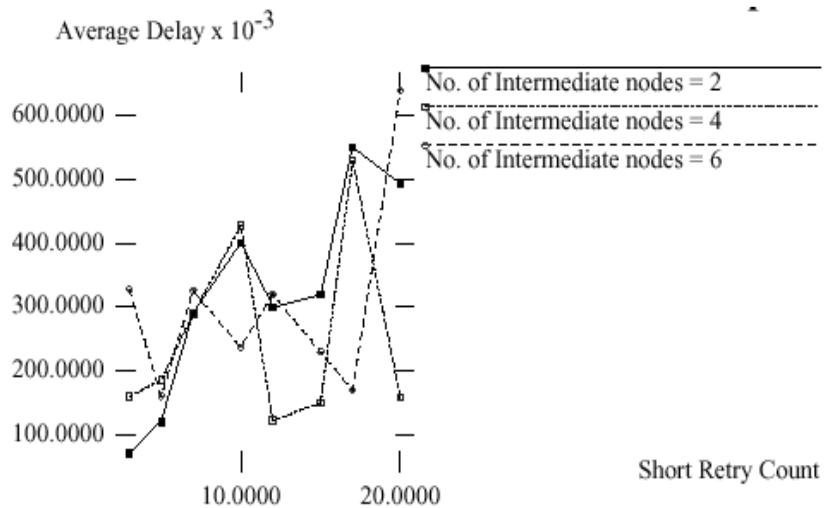

**Figure 3: Variation of Average End-to-End Delay with Short Retry Count**

**Short Explanation**: *The change in the value of Average delay is not uniform. Obviously as the number of retransmission attempts (short retry count) increases there is chance for increase in the Average Delay.*

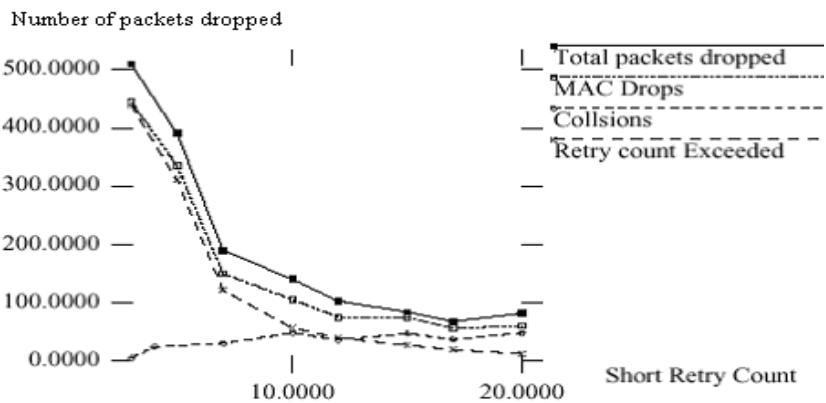

**Figure 4: Variation of Number of Packets Dropped with Short Retry Count**

**Short Explanation**: *Number of packets dropped decreases with the increase in the value of short retry count. At higher values of short retry count, node gets more chances for transmission of the packet, hence less number of packets are dropped.*

Short Retry Count, which is used for small sized packets, represents the number of retransmission attempts a node makes, before giving up and discarding a packet. In IEEE 802.11 its default value is 7. Lower values of short retry count are suitable when the number of transmission attempts by different nodes in the vicinity of a node is low. As the number of nodes attempting transmissions and retransmissions increases, lower values of short retry count result in deterioration in performance as shown in Fig. 2. As the number of common nodes and common links through which the packet has to pass through increases, higher values of short retry count are suitable.

When the value of short retry count is low the number of packets dropped due to retry count exceeding the limit is more than that of number of packets dropped due to collisions. The results are shown in Fig. 4. No fixed pattern is observed regarding the variation of average End-to-End delay with short retry count as in Fig. 3.

## The effect of CWMin and CWMax

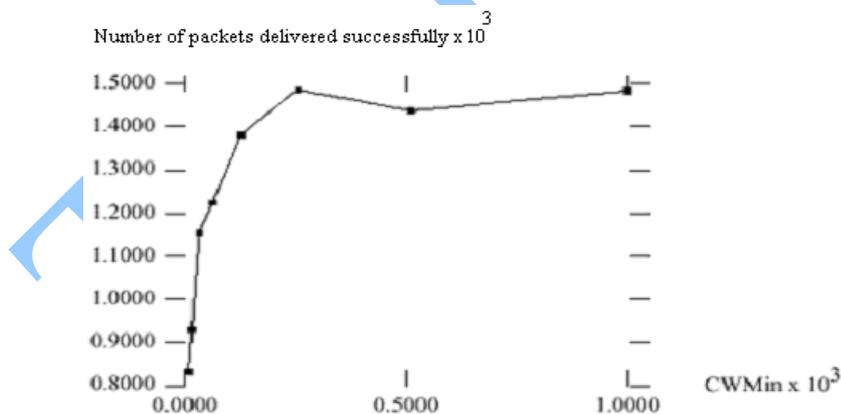

**Figure 5: Variation of Number of packets delivered successfully with CWMin**

**Short Explanation**: *Number of packets delivered increases with the increase in the value of CWMin. Number of packets delivered is low when CWMin value is low. The value of number of packets delivered successfully almost remains constant beyond some limit i.e. CWMin = 256.*

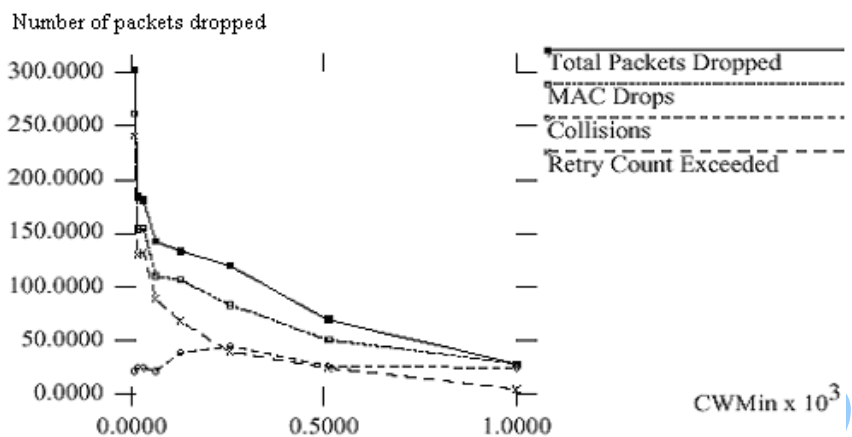

**Figure 6: Variation of Number of Packets Dropped with CWMin**

**Short Explanation**: *Number of packets dropped decreases with the increase in the value of CWMin. Low values of CWMin results in dropping of more number of packets.*

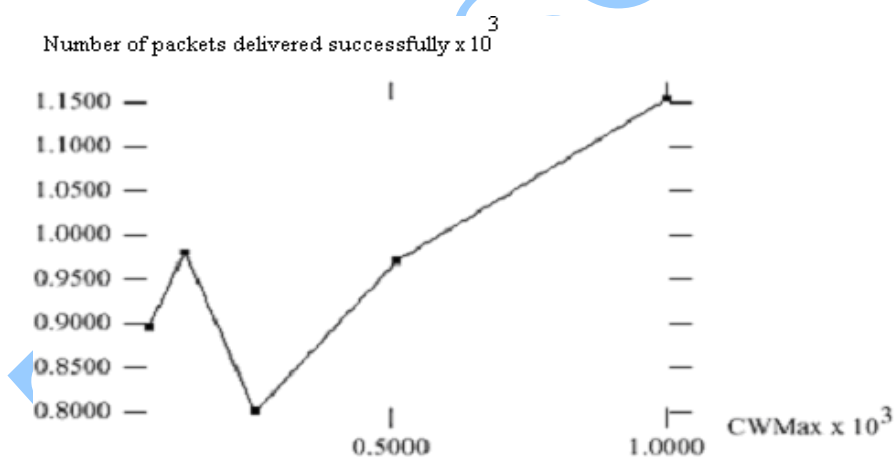

**Figure 7: Variation of Number of packets delivered successfully with CWMax**

**Short Explanation**: *Number of packets delivered successfully is better at higher values of CWMax, but the change is not uniform similar to CWMin.*

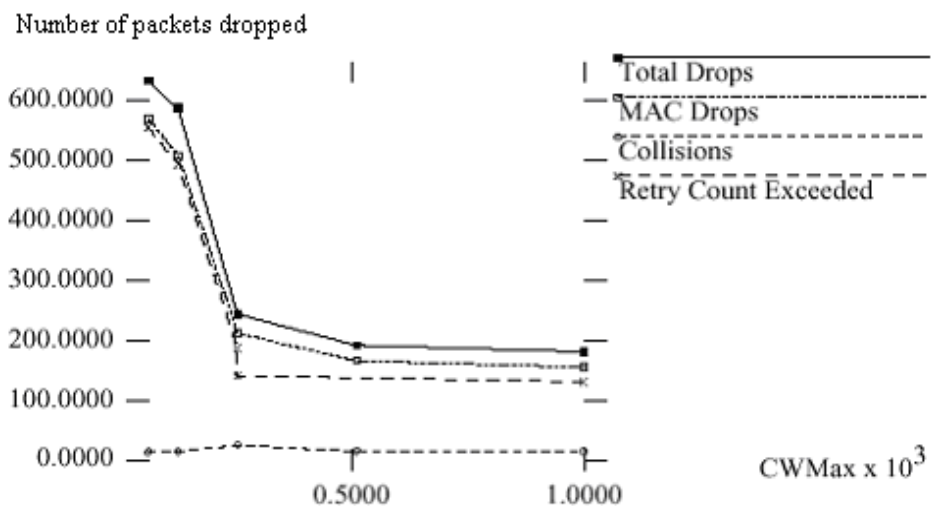

**Figure 8: Variation of Number of Packets Dropped with CWMax**

**Short Explanation**: *Number of packets dropped decreases with the increase in the value of CWMax. But the value remains almost constant when CWMax = 256.*

**Table 2: IEEE 802.11 with Different CW Limits**

| CW Limits (CWMin, CWMax) | Number of packets delivered successfully | Average End-to-End Delay (sec) | Total number of Packets Dropped | Number of Packets Dropped due to MAC Problems | Number of Packets Dropped due to Collisions | Number of Packets Dropped due to Retry Count Exceeding its Limit |
|---|---|---|---|---|---|---|
| (15,1023) | 1382 | 0.019914 | 226 | 180 | 51 | 129 |
| (31,1023) | 1309 | 0.08724 | 243 | 192 | 60 | 132 |
| (31,511) | 1266 | 0.17015 | 195 | 156 | 55 | 101 |
| (127,255) | 1423 | 0.02022 | 151 | 125 | 55 | 70 |
| (127,511) | 1598 | 0.31557 | 149 | 106 | 60 | 46 |
| (127,1023) | 1631 | 0.1462 | 165 | 122 | 68 | 54 |
| (255,511) | 1668 | 0.42972 | 93 | 73 | 54 | 19 |

**Short Explanation**: *Best values of (CWMin, CWMax) are (255,511). It is better to start with medium values of CWMin and CWMax.*

In IEEE 802.11 when a transmission attempt is successful, the value of Contention window size, CW is set to CWMin. This results in a node

181

selecting lower value for Backoff-interval. When a node becomes ready it sends an RTS. Similarly other nodes may select lower backoff-interval and transmit RTS resulting in RTS-RTS collision. When the value of CWMin is low the probability of RTS-RTS and RTS-CTS collisions is high resulting in packets being dropped due to retry count exceeding its limit. The solution for this is higher values of CWMin. The improvement in performance is observed with increase in CWMin value as shown in Fig. 5. The number of packets dropped decreases with increase in CWMin value, which is shown in Fig. 6. Similar is the case with CWMax. Higher values of CWMax result in better performance as shown in Fig. 7 and Fig. 8.

Interesting observation is, the impact of CWMin is more than that of CWMax. Hence CWMin has to be set carefully for better performance. In many situations higher number of packets delivered successfully is obtained by varying CWMin and CWMax slowly. The results in Table 2 are obtained when CW toggles between CWMin and CWMax.

**Conclusion**

The optimum values for various IEEE 802.11 parameters are determined. Higher values of short retry count are suitable for adhoc networks. Number of packets delivered successfully increases with the increase of short retry count and the optimum value of short retry count is 20. The optimum performance in terms of number of packets dropped is observed at short retry count 20. Slow increase and decrease of CW limits i.e., CWMin and CWMax result in improvement in performance of TCP over IEEE 802.11. CWMax value has little influence on the performance of TCP over IEEE 802.11. Number of packets delivered successfully increases and number of packets dropped decreases with increase in CWMin value. The optimum value of CWMin value is 255, independent of CWMax value.